# Information-based Preprocessing of PLC Data for Automatic Behavior Modeling

Brandon K. Sai[a,*], Jonas Gram[a], Thomas Bauernhansl[a,b]

[a]*Fraunhofer Institute for Manufacturing Engineering and Automation IPA, Nobelst. 12, 70569 Stuttgart, Germany*
[b]*Institute of Industrial Manufacturing and Management, IFF, University of Stuttgart, Germany*

* Corresponding author. Tel.: +49 711 970-1918; *E-mail address:* brandon.sai@ipa.fraunhofer.de

**Abstract**

Cyber-physical systems (CPS) offer immense optimization potential for manufacturing processes through the availability of multivariate time series data of actors and sensors. Based on automated analysis software, the deployment of adaptive and responsive measures is possible for time series data. Due to the complex and dynamic nature of modern manufacturing, analysis and modeling often cannot be entirely automated. Even machine- or deep learning approaches often depend on a priori expert knowledge and labelling. In this paper, an information-based data preprocessing approach is proposed. By applying statistical methods including variance and correlation analysis, an approximation of the sampling rate in event-based systems and the utilization of spectral analysis, knowledge about the underlying manufacturing processes can be gained prior to modeling. The paper presents, how statistical analysis enables the pruning of a dataset's least important features and how the sampling rate approximation approach sets the base for further data analysis and modeling. The data's underlying periodicity, originating from the cyclic nature of an automated manufacturing process, will be detected by utilizing the fast Fourier transform. This information-based preprocessing method will then be validated for process time series data of cyber-physical systems' programmable logic controllers (PLC).

*Keywords:* Data Preprocessing; Behavioral Modeling; OEE Optimization

**1. Introduction**

Cyber-physical systems (CPS) have the potential to revolutionize industrial manufacturing processes through the collection of multivariate time series data from actors and sensors. However, the complex and dynamic nature of modern manufacturing poses a significant challenge to automated analysis and modeling of the data. In this paper, we propose an information-based data preprocessing approach that aims to improve the accuracy and efficiency of automated behavior modeling for industrial manufacturing processes. The proposed approach employs statistical methods such as variance and correlation analysis to prune the dataset to retain only the most important features. This data size reduction makes it possible to gain a deeper understanding of underlying production processes. Additionally, we utilize the fast Fourier transform (FFT) to classify the data into real origin intervals or process cycles. This allows identifying of underlying periodic patterns in the data that originate from the cyclical nature of automated production processes. The proposed information-based preprocessing method will be validated using process time series data from a CPS' programmable logic controllers (PLC). Through this validation, we aim to demonstrate the effectiveness of the proposed approach in providing a deeper understanding of underlying production processes and its ability to improve the accuracy and efficiency of automated behavior modeling for industrial manufacturing processes.

## 2. Related work

Much research has been carried out in time series data preprocessing. In the following, some of the most important approaches relating to this papers main goal are shown. The focus is set on feature selection through variance and correlation-based methods and time series periodicity detection.

### 2.1. Time series data processing

Various ways for preprocessing time series data with regard to machine learning have been assessed, including normalization, scaling and detrending of time series data [1]. McGovern applies moving averages and frequency domain processing via Fourier transform to time series data [2]. Panja et al. conduct research for choosing a fraction of data points for a regression model without losing any features or trends in the data. This involves dividing the dataset along the axis into equal parts, along a specific curve or at certain change points [3]. Further research is carried out in preprocessing of web usage or Internet of Things (IoT) data, with a strong focus on the processing queues and pipelines. Additionally, real time processing and denoising of the data are evaluated [4, 5].

### 2.2. Feature selection with variance-based methods

Yan and Tang conduct research for feature screening by utilizing a data-slicing technique to construct a new index called the fused mean-variance. They use a mathematical approach to measure feature dependencies and importance [6]. Insolia et al. developed a feature selection and outlier detection method for regression problems. Feature selection is performed through a robust class of non-concave penalization methods estimating the importance of each data point. While being robust, this method relies on manual parameter fine-tuning [7]. Kamalov proposes an approach for feature selection based on orthogonal variance. First decomposing model output variance into orthogonal components based on feature subsets, then making use of variance decomposition to evaluate features [8]. Sadeghyan proposes feature selection through Sensitivity Analysis (SA) based on the work of Lauret et al. [9], where a ranking system is provided for all the existing features in order to remove the redundant features. A neural network is fitted to the features of the data set and probability density functions are assigned with the ranges of variation for each feature. The features' relevance is determined by analyzing the Fourier decomposition of the model output variance [10]. Nuisance features, which are at maximum loosely related to the underlying structure of the data have been identified utilizing the data graphs Laplacians eigenvectors as the Laplacian score [11].

### 2.3. Feature selection with correlation analysis methods

Extensive research has been performed regarding feature selection based on sparse methods like Cox regression and Elastic net regression modeling in various ways. Additionally, filter methods like random forest modeling for feature evaluation have been tested [12]. In the field of filter methods, sum of squared canonical correlation coefficients are used for feature ranking and a novel algorithm has been developed [13]. Making use of an AutoEncoder (AE) with a concrete layer [14], detecting correlated features has been a success. The AE is trained to reconstruct the time series data from a subset of selected features [11]. Furthermore, the Merit Score for time series data conducting feature subset selection based on the correlation patterns of single feature classifier outputs has been proposed. The score ranks feature importance if a good feature subset contains features highly correlated to the class but uncorrelated to each other [15]. Diez et al. evaluate certain linear and non-linear correlation methods for feature selection. This includes the linear absolute Pearson coefficient and the non-linear geometric mean of marginal entropies which is the mutual information of features normalized by the tightest bound [16]. Dewaskar et al. utilize iterative hypothesis testing to identify stable bimodules, which satisfy a natural stability condition. Bimodules are groups of features with significant aggregate cross-correlation and can be used for feature selection of time series data [17]. Further research is being performed in the field of Graph Neural Network (GNN) modeling to identify feature correlations. Feature correlation aggregation is proposed to learn the second order information from feature correlation between a node and its neighbors to enable improved feature generation by the GNN [18].

### 2.4. Time series periodicity detection

For periodicity detection, a sliding window approach is proposed by Długo et al. Per window subset, the homology and its L1 and L2 norms are computed and portioned into k equispaced points via k-means clustering or min-max transformation of the data [19]. Periodicity detection is performed based on an optimization algorithm of Dynamic Time Warping (DTW), Robust DTW and a time series slicing algorithm as well. This concept improves efficiency and robustness to noise and outliers for recognition of periodicity in time series [20]. Barnet et al. present a Gaussian Mixture periodicity detection algorithm iteratively using an Integral Convolution approach. The algorithm calculates a point-wise estimate of the explained data for a given integrated subset, which is tested as an underlying periodicity of the time series data [21]. In terms of statistical detection of periodic signals, the Bayesian- and the frequentist period search have been evaluated. The Bayesian approach computed the Bayes factor based on marginal likelihood to determine whether a feature is periodic, while the frequentist approach relies on hypothesis testing based on the likelihood ratio, computed through ratios of likelihood maxima [22]. Periodicity detection is also achieved by utilizing the multifrequency periodogram based on least-square regression and hypothesis testing. This is optimized for efficiency, parallelism and robustness against noise in the data [23]. Covino et al. conduct periodicity detection based on Gaussian processes [24, 25]. These processes are generative models and an infinite dimensional extension of the multivariate Normal distribution. A Bayesian nonparametric approach is provided, where the mean is set to none, a linear or quadratic function and the covariance is a positive semidefinite function. The covariance kernel is chosen based on a priori knowledge of the time series data. Further research is performed in the identification of periodic signals through the Auto Correlation function [26, 27]. An automated period detection method has been developed relying on the discrete Fourier transform. The time series data is transformed into the frequency domain, and the frequency domain probability density function is solved to recognize periodicity. This function depends on amplitude, frequency, phase and least-squares fitting of trigonometric functions to the transformed data [28–30]. In addition to uniform Fourier transforms, non-uniform fast Fourier transforms for unevenly sampled data have been proposed [31–34]. Another advanced technique based on the Fourier transform is the discrete Double Fourier transform for the detection of periodic patterns [35]. Resting upon the developments on the Fourier transform, period detection can be conducted through Fourier-likelihood periodograms. This is achieved by converting the time series in the frequency domain via the Fourier transform, computing the periodogram [36], converting the values for each sample to Gaussian likelihoods and taking the product of these across all data points [37]. Human sleep cycle detection has been done using cosine functions and a cosinor model [38] to describe pseudo-periodical patterns [39].

## 3. Contributions

The information-based preprocessing approach of this paper, can be applied after standard processing of PLC time series data, like removing rows and columns with empty values and correctly naming the columns. PLC data consist of discrete values from sensors and actors of the underlying CPS. Each data source (sensor or actor) is referred to as feature in the time series data, represented as a column. Additionally, these sources will be referred to as signals in the PLC data stream.

Table 1: Model dataset for PLC time series data.

| Timestamp (h:mm:ms) | Actuator | Sensor 1 | Sensor 2 | Sensor 3 |
|---|---|---|---|---|
| 3:10:000 | 0 | 0 | 0 | 0 |
| 3:10:500 | 0 | 1 | 0 | 0 |
| 3:11:000 | 0 | 1 | 0 | 0 |
| 3:12:000 | 1 | 1 | 0 | 0 |

This approach aims to enable methods in generating process understanding through modeling or analyzing by removing the least important PLC signals and dividing the dataset into the periodic production cycles. The proposed preprocessing method generates a-priori knowledge of the CPS without the need for extensive modeling or expert knowledge.

### 3.1. Automated feature selection

Automated production equipment in general entails multiple sensors and control-parameters. When logging the generated data from the PLC, many of these signals are not necessary to gain information about the underlying production process. The goal is to filter the signals with a high probability of having a marginal impact on the eventual modeling or analysis of the data. Since many PLC signals are only used occasionally or not at all, a variance threshold per signal is proposed. The PLC time series data is described as $X = (x_1, \ldots, x_n) \in \mathbb{R}^{n \times d}$ with $n$ samples and dimensionality $d$. The dataset consists of $d$ feature columns $c = (c_1, \ldots, c_d)$, where $c_k$ is the $k$-th column of the dataset $X$. In modern complex automated production sites, $d$ will be a high number that this approach tries to diminish prior to modeling or further usage of the data. The variance $\sigma^2$, the squared deviation of a value from the mean, is calculated per feature column $c = (a_1, \ldots, a_n) \in \mathbb{R}^{n \times 1}$, where $a_i$ is the $i$-th element of a column.

$$\sigma^2 = \frac{\sum_{i=1}^{n}(a_i - \mu)^2}{n} \quad (1)$$

Each feature with a variance lower than a certain threshold $thr$ is subsequently removed, since a low variance reflects a menial rate of value alterations. Complementary, to further reduce the dimensionality of the dataset, highly correlated features will be removed. Through the utilization of correlation analysis, redundant columns, that have low or no impact on modeling or the analysis of the data, will be detected [40]. The Spearman correlation coefficient $R_{km}$ is calculated to determine the pairwise feature correlation of each dataset column $c_k$ with every other column $c_m$. If the modulus of the correlation between two features exceeds a certain threshold, one of the features is subsequently removed. First, the rank $r_i$ per element is computed for each column. Per column, the lowest element is assigned the rank $r_0 = 0$, while the highest is assigned $r_n = i$. In the case of a value tie, each of the tied elements is assigned the average of the ranks that would have been assigned to all the tied values. Then the Spearman correlation is computed for each pair of features with n elements and the rank value $r_{ik}$ for each.

$$R_{km} = \frac{\frac{1}{n}\sum_{i=1}^{n}(r_{ik} - \overline{r_k})(r_{im} - \overline{r_m})}{\sqrt{\frac{\sum_{i=1}^{n}(r_{ik} - \overline{r_k})^2}{n}} * \sqrt{\frac{\sum_{i=1}^{n}(r_{im} - \overline{r_m})^2}{n}}} \quad (2)$$

Both described methods ensure a faster data handling due to dimensionality reduction without the loss of crucial information.

## 3.2. Resampling the event-based PLC time series data

Due to the PLC having a maximum data rate for the connection interfaces, PLC time series data is often generated event based. Even though the data rate might be sufficient for data sampling at a higher frequency, in production environments, it is important not to overload the PLC with secondary tasks due to failures having a high impact on costs. Because modern supervisory control and data acquisition (SCADA) systems often realize uniform data logging, this step is not necessary for each production system. In the process of event-based time series data sampling, the next dataset row consisting of one value per PLC signal is generated whenever one of the signal values just changes. As a result, the dataset is non-equidistant or non-uniform because the time delta t between the event-based generated timestamps of the rows varies. Many analyzing or modeling methods work with non-uniform data, but some will not, for example, the cycle detection method proposed in chapter 3.3. In order to enable data analysis and modeling approaches depending on uniform data, a simple upscaling method for event-based time series data X is proposed, to create the uniform time series dataset $\hat{X} = (\hat{x}_1, \ldots, \hat{x}_{\hat{n}}) \in \mathbb{R}^{\hat{n} \times d}$. In the uniform time series dataset, the time interval between each row will be the same. The new timedelta $\hat{t}$ for $\hat{X}$ describing the time amount between two measurements is set to the PLC cycle time of the data generating PLC. This leads to the uniform dataset measuring frequency $\hat{f}$ of how many rows of data are generated per second.

$$\hat{f} = \frac{1000}{\hat{t}} \quad (3)$$

Now the dataset $\hat{X}$ is resampled starting from row $x_1$ which is the same in X and $\hat{X}$. Until the final timestamp of $x_n$ is reached, a new row $\hat{x}_i$ is generated. Due to the logic of the event-based data, every row $\hat{x}_{i+1}$ matches the values of $\hat{x}_i$ until the next event-timestamp at $x_i$ is reached and the values change to the of $x_i$.

## 3.3. Periodic cycle time detection

Based on the uniform time series dataset $\hat{X}$, a robust production process cycle time detection method is developed. The goal is to gain information about the distinction and duration of the periodic production cycles to support further analysis prior to or after modeling the data. First, each column $\hat{c}_{\hat{k}}$ of the uniformed PLC time series $\hat{X}$ is transformed into frequency domain $F_k$ by utilizing the 1-dimensional discrete uniform fast Fourier transform (FFT) [41]. The algorithm computes every occurring periodic frequency in the time series data as waves and maps it together with its importance amplitude, such that $F_k = (f_1, \ldots, f_{\hat{n}}) \in \mathbb{R}^{\hat{n} \times 2}$. Each element $f_i$ consists of a frequency value in $Hz$ depicting the detected periodicity and an amplitude value to illustrate the importance of this periodic frequency for the dataset. In order to analyze the most important periodicity of the dataset, which is the production cycle time, a further analyzing method is proposed.

Per column, each value with an amplitude below the top 30% is discarded. Additionally, a certain percentage of frequency values starting from zero is being treated as anomalies and discarded as well. This is based on the assumption that the underlying production process will have anomaly extended cycle times during periods of maintenance, setup-time or holidays. Now the local maxima of the remaining amplitudes are identified and their related data samples collected. Based on the FFT and this algorithm, the cycle time $T$ of the PLC time series datasets production process is the inverse of the frequency with the corresponding highest amplitude. Utilizing this information, the responsible signal for the strongest frequency can be identified. This signals pattern together with the cycle time can be used to partition the PLC time series into the respective production cycles.

## 4. Application and validation

To validate the proposed method and present possible applications, the method has been applied to a synthetic and a real PLC time series dataset.

### 4.1. Synthetic dataset

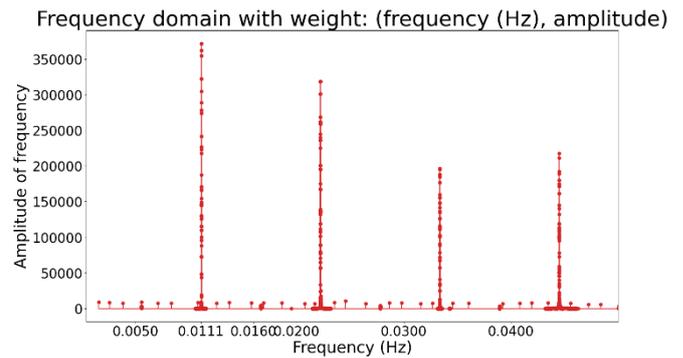

Figure 1: Frequency analysis of the synthetic dataset. The x-axis shows occurring frequencies in the data, the y-axis measures the frequencies importance via the amplitude.

To simulate a controlled environment in order to validate the proposed information-based preprocessing approach, an event-based synthetic dataset has been generated. This time series dataset consists of 35 features over a week of simulated production cycles and an optimum cycle time of 90 seconds. Each cycle consists of 17 different states. The dataset was generated with 10% noise. In this context, noise is the presence of acyclic or infrequent activities in the sensor and actor data. By applying the automated feature selection, one signal was removed due to low variance, and one was removed due to high correlation with other signals. Then the dataset was resampled by upscaling it to match $20ms$ of time between measurements. Subsequently, the cycle time detection was realized. The software could detect the optimum cycle time of 90 seconds and no information was lost by the automated feature selection. Additionally, the strongest cyclic signal was identified as signal 25, which can now further be used to partition the dataset into production cycles.

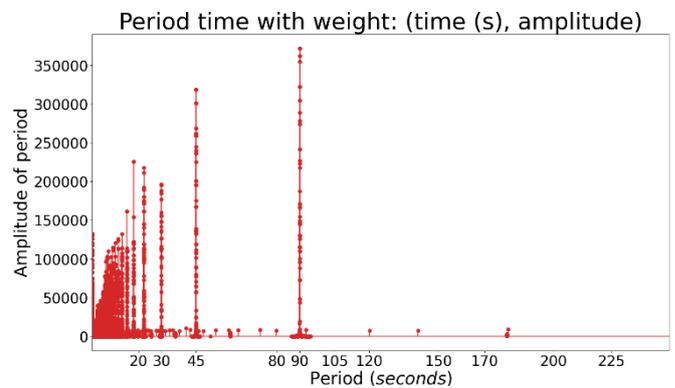

Figure 2: Transformation of the frequency analysis into the time domain. The strongest detected period was 90 seconds, visible at the highest amplitude point.

*4.2. Real PLC dataset*

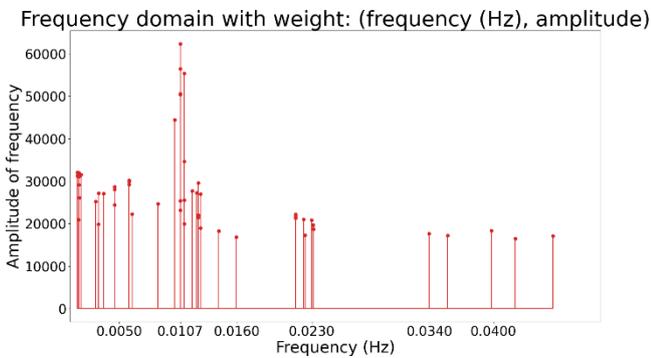

Figure 3: Frequency analysis of the real-world production dataset. The x-axis shows occurring frequencies in the data, the y-axis measures the frequencies importance via the amplitude.

For true validation of the preprocessing method, the approach was tested in a real CPS production environment.

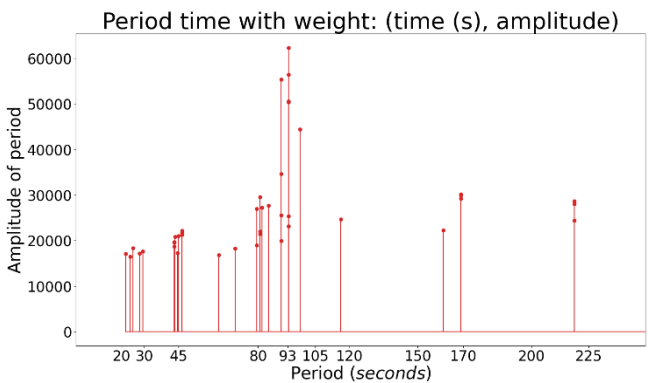

Figure 4: Transformation of the frequency analysis into the time domain. The strongest detected period was about 93 seconds, visible at the highest amplitude point.

This time series dataset has been recorded event-based as well. It consists of 25 features, logged over one week. The optimum cycle time is 93 seconds with each cycle consisting of 14 different states. Using the automated feature selection, 12 PLC signals were dropped out of the dataset for low variant behavior. Another signal has been dropped due to high correlation, reducing the data dimensionality to 12 features. To match the uniform PLC sampling, the data has then been resampled to match 10ms of time between measurements. The cycle detection approach identified the optimum cycle time of about 93s. The software identified the strongest cyclic signal as either signal 5 or signal 10.

Even though both signals show the same strong cyclic behavior, in reality, only signal 5 is used as the cycle starting signal in the PLC software.

## 5. Conclusion

This paper proposes a method for preprocessing cyber-physical systems time series data. The approach automatically generates the needed information regarding the underlying process and dataset, which otherwise would have been manually analyzed by experts. The processed data enables modeling or simulation approaches due to lower dimensionality and insight into the exact periodic behavior. The method shows good capability of time series data dimensionality reduction and reliably detects periodic production cycle times. Future work may improve the detection of the strongest cyclic signal to ensure an optimized dataset partitioning into cycles. Additionally, the data resampling approach presented in 4.2 can be automated as well. Further automation potential is to dynamically and automatically set the threshold values for the feature-selection and the percentage of discarded values during the cycle time detection. This information-based preprocessing approach will be continuously applied and optimized in the future.

## Appendix A. Additional outputs for the production dataset validation

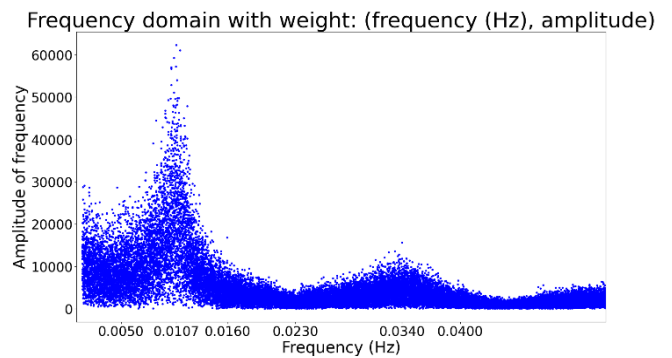

Figure 5: Distribution of the importance (y-axis) of the frequency spectrum (x-axis) for signal 5.

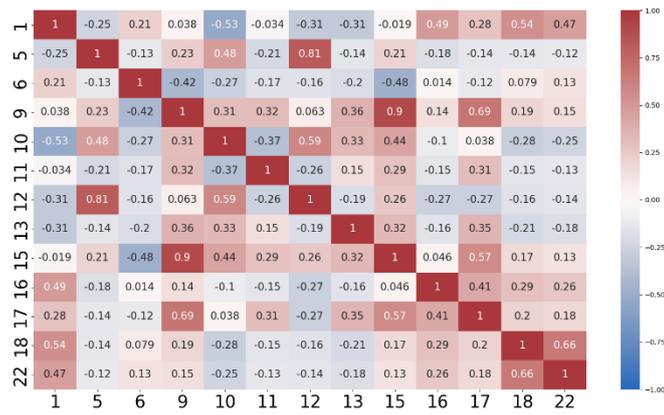

Figure 6: Feature correlation matrix of the production dataset after removal of the low variance features.